\documentclass[lettersize,journal]{IEEEtran}
\pdfoutput=1
\usepackage{amsmath,amsfonts}
\usepackage{algorithmic}
\usepackage{algorithm}
\usepackage{array}
\usepackage{textcomp}
\usepackage{stfloats}
\usepackage{url}
\usepackage{verbatim}
\usepackage{graphicx}
\usepackage{bm}
\usepackage{multirow}
\usepackage{multicol}
\usepackage{enumitem}
\usepackage[table,xcdraw]{xcolor}

\usepackage[frozencache]{minted}
\definecolor{LightGray}{gray}{0.9}

\newcolumntype{L}[1]{>{\raggedright\arraybackslash}m{#1}}
\newcolumntype{C}[1]{>{\centering\arraybackslash}m{#1}}
\newcolumntype{R}[1]{>{\raggedleft\arraybackslash}m{#1}}

\makeatletter
\newcommand\notsotiny{\@setfontsize\notsotiny\@vipt\@viipt}
\makeatother

\usepackage[binary-units=true]{siunitx}

\DeclareSIUnit{\fps}{fps}
\DeclareSIUnit{\loc}{LoC}

\makeatletter
\let\MYcaption\@makecaption
\makeatother
\usepackage[font=footnotesize]{subcaption}
\makeatletter
\let\@makecaption\MYcaption
\makeatother

\DeclareCaptionLabelSeparator{periodspace}{.\quad}
\usepackage{enotez}
\usepackage{setspace}
\setenotez{counter-format=symbols}
\setenotez{list-name=Notices}
\DeclareInstance{enotez-list}{custom}{paragraph}
{
	heading = \section*{#1} ,
	notes-sep = 0.2\baselineskip ,
	format = \footnotesize ,
	number = \textsuperscript{#1}
}

\usepackage{scalerel}
\usepackage{tikz}
\usetikzlibrary{svg.path}

\definecolor{orcidlogocol}{HTML}{A6CE39}
\tikzset{
	orcidlogo/.pic={
		\fill[orcidlogocol] svg{M256,128c0,70.7-57.3,128-128,128C57.3,256,0,198.7,0,128C0,57.3,57.3,0,128,0C198.7,0,256,57.3,256,128z};
		\fill[white] svg{M86.3,186.2H70.9V79.1h15.4v48.4V186.2z}
		svg{M108.9,79.1h41.6c39.6,0,57,28.3,57,53.6c0,27.5-21.5,53.6-56.8,53.6h-41.8V79.1z M124.3,172.4h24.5c34.9,0,42.9-26.5,42.9-39.7c0-21.5-13.7-39.7-43.7-39.7h-23.7V172.4z}
		svg{M88.7,56.8c0,5.5-4.5,10.1-10.1,10.1c-5.6,0-10.1-4.6-10.1-10.1c0-5.6,4.5-10.1,10.1-10.1C84.2,46.7,88.7,51.3,88.7,56.8z};
	}
}

\newcommand\orcidicon[1]{\hspace*{1px}\textsuperscript{\footnotesize \href{https://orcid.org/#1}{\mbox{\scalerel*{
					\begin{tikzpicture}[yscale=-1,transform shape]
						\pic{orcidlogo};
				\end{tikzpicture}}{d}}}}}

\usepackage[
style=ieee,
backend=biber,
maxcitenames=1,
mincitenames=1,
uniquelist=false,
maxbibnames=999,
maxnames=1,
minnames=1,
doi=true,
url=false,
isbn=false,
eprint=false
]{biblatex}

\AtBeginBibliography{\scriptsize}

\DeclareSourcemap{
	\maps[datatype=bibtex]{
		\map[overwrite=true]{
			\step[fieldset=abstract, null]
		}
	}
}
\DeclareSourcemap{
	\maps[datatype=bibtex]{
		\map[overwrite=true]{
			\step[fieldset=comment, null]
		}
	}
}

\AtEveryBibitem{%
	\clearfield{location}%
	\clearfield{volume}%
	\clearfield{pages}%
	\clearfield{number}%
}
\DeclareSourcemap{
	\maps[datatype=bibtex]{
		\map[overwrite=true]{
			\step[fieldset=organization, null]
		}
	}
}

\usepackage{xspace}
\newcommand{\dirquote}[1]{``#1''}

\newcommand{\eigName}[1]{``#1''}

\makeatletter
\long\def\@IEEEtitleabstractindextextbox#1{\parbox{0.966\textwidth}{#1}}
\makeatother

\newcounter{question}
\newcommand{\questionsec}[1]{
	\refstepcounter{question}
	\subsection*{\textbf{Q\arabic{question}) #1}}
}

\usepackage[%
textsize=footnotesize
]{todonotes}
\usepackage{hyperref}
\AtBeginDocument{
	\hypersetup{
		pdftitle = {GPU Performance Portability Needs Autotuning -- Demonstrating portable and performant cross-platform LLM kernels},
		pdfauthor = {Burkhard Ringlein, Tom Parnell, Radu Stoica},
	} 
}

\usepackage[acronym]{glossaries}
\glsdisablehyper

\begin{document}

\newacronym{vm}{VM}{virtual machine}
\newacronym{ml}{ML}{machine learning}
\newacronym{nn}{NN}{neural network}
\newacronym{llm}{LLM}{Large Language Models}
\newacronym{sla}{SLA}{service level agreement}
\newacronym{ig}{IG}{information gain}
\newacronym{bo}{BO}{Bayesian Optimization}
\newacronym{bbo}{BBO}{black box optimization}
\newacronym{gemm}{GEMM}{General Matrix Multiply}
\newacronym{ir}{IR}{Intermediate Representation}
\newacronym{loc}{LoC}{Lines of Code}
\newacronym{ai}{AI}{Artificial Intelligence}
\newacronym{dsl}{DSL}{Domain-specific Language}
\newacronym{cisc}{CISC}{Complex Instruction Set Computer}
\newacronym{risc}{RISC}{Reduced Insturction Set Computer}
\newacronym{jit}{JIT}{Just-in-Time}
\newacronym{sota}{SOTA}{state-of-the-art}

\title{GPU Performance Portability Needs Autotuning\\\Large{Demonstrating portable and performant cross-platform LLM kernels}}

\author{
	Burkhard~Ringlein\orcidicon{0000-0002-7222-9539}, 
	Thomas~Parnell\orcidicon{0000-0002-1308-6590}, 
	and Radu~Stoica\orcidicon{0009-0005-8089-866X} 

\thanks{B. Ringlein, T. Parnell and R. Stoica are with IBM Research Europe, Säumerstrasse 4, 8803 Rüschlikon, Switzerland. E-mail: \{ngl, tpa, rst\}@zurich.ibm.com }
}

\markboth{B. Ringlein \MakeLowercase{\textit{et al.}}: GPU Performance Portability needs Autotuning}{B. Ringlein \MakeLowercase{\textit{et al.}}: GPU Performance Portability needs Autotuning}


\maketitle
\begin{abstract}

As LLMs grow in complexity, achieving state-of-the-art performance requires tight co-design across algorithms, software, and hardware. Today’s reliance on a single dominant platform limits portability, creates vendor lock-in, and raises barriers for new AI hardware. In this work, we make the case for combining just-in-time (JIT) compilation with 
comprehensive
kernel parameter autotuning to enable portable LLM inference with state-of-the-art performance without code changes. 
Focusing on performance-critical LLM kernels, 
we demonstrate that this approach explores up to $15\times$ more kernel parameter configurations, produces significantly more diverse code across multiple dimensions, and even outperforms vendor-optimized implementations by up to 230\%, all while reducing kernel code size by $70\times$ and eliminating manual code optimizations. 
Our results highlight autotuning as a promising path to unlocking model portability across GPU vendors.



\end{abstract}

\begin{IEEEkeywords}
Language Models, Portability, Domain-specific Languages, Performance of Systems, Code tuning
\end{IEEEkeywords}

\section{Introduction}

\glspl{llm} have evolved dramatically in the past years. Besides the improvement in model architectures and training procedures, there have been many innovations in optimizing LLM applications for modern hardware (\cite{ DaoFlashAttention2FasterAttention2023, ShahFlashAttention3FastAccurate2024, KwonEfficientMemoryManagement2023, YeFlashInferEfficientCustomizable2025}).
However, this race in features and performance leads to a \dirquote{hardware lottery} \cite{Hooker2021} for new \gls{ai} or \gls{ml} paradigms and to a gravity slope around the most dominant hardware platform. The tight interconnect between AI algorithms and AI hardware leads to limitations on the deployment and application scenario of \gls{ai}, since most features are only supported for a narrow set of hardware or input problem sizes~\cite{Hooker2021}. 
Consequently, the number and the size of libraries used to deploy \glspl{llm} with \gls{sota} performance have grown dramatically. 

\begin{figure}[t]
	\centering
	\hspace*{-2.5ex}
	\begin{subfigure}[t]{.19\textwidth}
		\includegraphics[width=\linewidth]{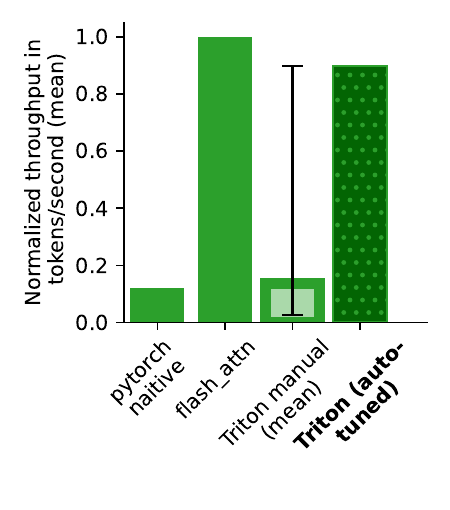}
		\vspace*{-2.5em}
		\caption{A100}
		\label{subfig:mot-a100-perf}
	\end{subfigure}%
	\hspace*{-1.5ex}
	\begin{subfigure}[t]{.19\textwidth}
		\includegraphics[width=\linewidth]{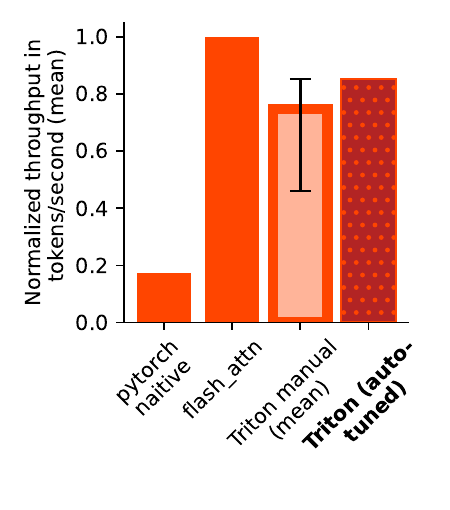}
		\vspace*{-2.5em}
		\caption{MI250}
		\label{subfig:mot-mi250-perf}
	\end{subfigure}%
	\hspace*{-2ex}
	\begin{subfigure}[t]{.19\textwidth}
		\includegraphics[width=\linewidth]{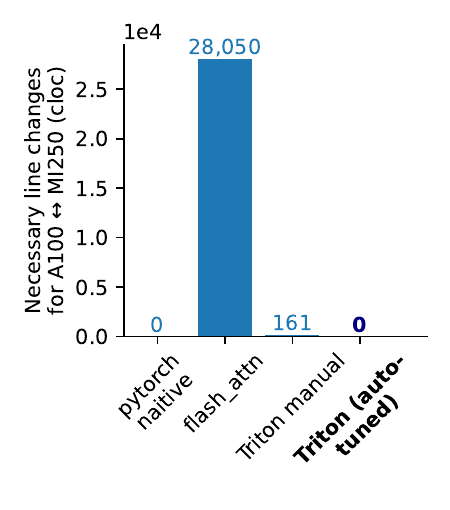}
		\vspace*{-2.5em}
		\caption{\texttt{cloc diff}}
		\label{subfig:mot-line-diff}
	\end{subfigure}%
	\vspace*{-0.5em}
	\caption{Normalized throughput (a, b) and the effort to port the attention layer across GPU architectures (c). Workload: Attention layer for Llama3.1-8b, batch size 64, sequence length 1024. 
	}
	\label{fig:motivational-example}
	\vspace*{-1em}
\end{figure}

We highlight this dynamic in Fig. 1, where the performance of four different implementations of the core flash-attention layer~\cite{VaswaniAttentionAllYou2017} is shown on two GPU architectures from different vendors. The flash attention implementations are listed in \autoref{tab:impl-used}. The performance results are normalized to the baseline PyTorch native implementation on each platform. 
As can be seen, the generic native PyTorch implementation requires only $29$ \gls{loc} but is $6-13\times$ slower than the popular \texttt{flash\_attn} library optimized for NVIDIA GPUs or the \textit{ROCm} version of the flash attention library offering \gls{sota} performance on the AMD MI250. However, the two optimized libraries are also significantly more complex than the PyTorch native implementation (\SI{2300}{\times} more LoC).
Finally, \autoref{subfig:mot-line-diff} quantifies the low-level code changes required to port the LLM attention layer between the NVIDIA and AMD architectures. To achieve \gls{sota} performance on the MI250, more than \SI{40}{\percent} of the initial \textit{flash\_attn} had to be manually optimized.

Having a kernel code that has the conciseness and portability of PyTorch but also SOTA performance is still an open research question.
Writing tens of thousands of \gls{loc} to port a one-line kernel~\cite{VaswaniAttentionAllYou2017} slows down research, complicates deployment of new \gls{ml} methods significantly, and hinders adoption of new hardware. Additionally, orders of magnitude larger code also means a proportionally higher probability of making mistakes. 
These problems are even more pressing, since none of the aforementioned attention libraries are \eigName{final}. Changes are constantly required to incorporate new algorithms, requirements, or support for new hardware (\cite{Hooker2021, KwonEfficientMemoryManagement2023}).
For example, it took over a year to adapt the \texttt{flash\_attn} library to the new NVIDIA Hopper architecture~\cite{DaoFlashAttention2FasterAttention2023, ShahFlashAttention3FastAccurate2024}

In this work, we revisit the question of zero-change performance-portability, aiming to achieve concise yet portable and efficient GPU kernels.
Previous work has shown that kernel autotuning, with its ability to search the space of possible kernel configuration parameters and automatically adapt to different architectures, would be a promising direction~\cite{Ringlein2022, Diamantopoulos2020b, Moreau2018}.
In this study, we provide the first evidence that autotuning can help achieve SOTA performance for two platform-independent kernel implementations on two different GPU platforms from two vendors for LLM deployments. To contrast with the earlier discussed motivational examples, we also show the corresponding results using our flash-attention autotuned solution in \autoref{fig:motivational-example}.
We then discuss the state-of-the-practice regarding autotuning and the underlying issues that prevent it from being used more widely. Finally, we highlight how future work could enhance the applicability of autotuning.



\section{Background and Related Work}

\subsection{Code Generation Based on Templated Libraries}
One popular approach to specializing kernels for different hardware architectures is through the use of template libraries.
For example, one set of popular template libraries that aims to implement the attention layer~\cite{VaswaniAttentionAllYou2017} as efficiently as possible and for a wide range of usage scenarios are \texttt{flash\_attn}~\cite{DaoFlashAttentionFastMemoryEfficient2022, DaoFlashAttention2FasterAttention2023, ShahFlashAttention3FastAccurate2024} and \texttt{FlashInfer}~\cite{YeFlashInferEfficientCustomizable2025}.
These libraries are non-trivial. In total, \texttt{flash\_attn} has nearly $\SI{70000}{\loc}$, while \texttt{Flashinfer} has $\SI{51000}{\loc}$.
These template libraries select which handwritten code fragments (written in a low-level language such as CUDA) to use based on the usage scenario (e.g., depending on the tensor shape, data type, or batch size). The low-level functions are then compiled to the corresponding accelerator instruction set. 

\clearpage
\vspace*{-1.52em}  
The template-based approach can produce point-wise optimal solutions and also cover a larger optimization space than a naïve approach. However, it is not perfect. 
First, the flexibility of such a template library is quite limited, since there are usually only one or a few ways to instantiate a template. 
This results in suboptimal code generation if a template is used on different hardware than the one on which the template was developed~\cite{Ringlein2022}. 
Suboptimal code generation could then lead to low compute utilization of GPUs for \gls{ai} applications~\cite{ShahFlashAttention3FastAccurate2024}.
Second, as hardware is constantly evolving (i.e., we now have tens of accelerator options based on different hardware architectures, with multiple generations and from a growing number of vendors), it makes choosing the right template version more difficult~\cite{Ringlein2022}. 
Third, adding a new template version optimized for a new set of accelerators requires significant manual effort -- measured in tens of thousands of \gls{loc} -- which makes the LLM software a high barrier to entry. 

\vspace*{-0.8em}
\subsection{Compilation vs. Autotuning}
\label{subsec:compilation-vs-autotuning}


Generally, there are three ways to avoid the need to write hand-optimized code: First, rely on ahead-of-time compilation to generate generic binaries. Second, use a \gls{jit} compiler to generate executables on the fly, taking execution-time data into account. And third, leverage autotuning to optimize for specific kernel usage scenarios.
Compiler-based approaches rely on predefined heuristics to generate general-purpose code. To improve performance, a compiler may employ multiple optimization passes and leverage a wide range of pre-determined heuristics. 
However, since the problem of compiling a high-level language to a lower-level language is NP-hard, optimizing performance can lead to very long compilation times~\cite{Ringlein2022}, which are not tolerable, especially for \gls{jit} compilers. 
Moreover, compilers must consider a wide range of possible kernel parameters to generate a valid binary. Therefore, they cannot maximize performance for all parameter combinations. 

In contrast, autotuned kernels augment compilation with empirical performance tuning, generating and benchmarking a wide range of kernel variants to select the best-performing configuration for the target hardware and scenario. 
Autotuning reduces the parameter space a compiler needs to consider for a specific kernel compilation. Therefore, it enables far better scenario-specific compilation optimizations, with the trade-off of having more compiled artifacts for each tuned target scenario.
This method can explore significantly more of the optimization space -- often an order of magnitude more variants~\cite{Diamantopoulos2020b, Moreau2018} -- leading to higher performance and better code specialization. 
While autotuning introduces additional overhead, its ability to deliver near-optimal performance without manual tuning makes it a compelling solution for deploying LLMs across heterogeneous platforms. 

These advantages make autotuning more suited for deploying LLMs on heterogeneous hardware platforms. Generally, autotuning must be balanced so that the performance advantages outweigh the disadvantages in terms of compilation and execution time overheads.

\vspace*{-1em}
\subsection{Triton: A tiling DSL}
\label{subsec:triton}
\label{sec:triton}

\begin{listing}[t]
	\captionof{listing}{A simple vector add program in Triton. 
	}
	\label{lst:triton-vector-add}
	\begin{minted}
		[
		%frame=lines,
		%framesep=2mm,
		%baselinestretch=1.2,
		%bgcolor=LightGray,
		fontsize=\scriptsize,
		%linenos
		]
		{python}
instance_id = tl.program_id(axis=0)
my_block_start = instance_id * BLOCK_SIZE
offsets = my_block_start + tl.arange(0, BLOCK_SIZE)
mem_mask = offsets < n_elements
x = tl.load(x_ptr + offsets, mask=mem_mask)
y = tl.load(y_ptr + offsets, mask=mem_mask)
result = x + y
tl.store(output_ptr + offsets, result, mask=mem_mask)
	\end{minted}
\end{listing}

\begin{table}[t!]
    \vspace*{-1em}
	\centering
	\caption{Investigated LLM kernel implementations
	}
	\label{tab:impl-used}
	{\notsotiny
\begin{tabular}{C{0.2ex}l|r|c|l}
\textbf{} &
\textbf{Implementation} &
\multicolumn{1}{c|}{\textbf{LoC}} &
\textbf{Target vendor} &
\textbf{Source} \\ \hline
\multirow{5}{*}{\hspace*{-2ex}\rotatebox[origin=c]{90}{\textbf{Attention}}} &
\texttt{flash\_attn} &
69197 &
NVIDIA &
\begin{tabular}[c]{@{}l@{}}\href{https://github.com/Dao-AILab/flash-attention}{github.com/Dao-AILab/flash-attention},\\[-0.1em] \cite{DaoFlashAttentionFastMemoryEfficient2022, DaoFlashAttention2FasterAttention2023}\end{tabular}
\\
&
\texttt{rocm\_flash\_attn} &
52489 &
AMD &
\href{https://github.com/ROCm/flash-attention}{github.com/ROCm/flash-attention} \\
&
pytorch native &
29 &
NVIDIA / AMD &
\href{https://github.com/pytorch/pytorch/blob/6055a4f612782ca944f2e0465f7497b7f18de4e9/torch/nn/functional.py\#L5732}{pytorch/.../functional.py},~\cite{VaswaniAttentionAllYou2017} \\
&
Triton manual &
1049 &
NVIDIA / AMD &
\cite{AMDTritonkernelsteamTriton_flash_attentionpy2024} \\ \cline{2-5} 
&
\textbf{Triton w/ autotuning} &
1100 &
NVIDIA / AMD &
\begin{tabular}[c]{@{}l@{}}\href{https://ibm.biz/vllm-ibm-triton-lib}{ibm.biz/vllm-ibm-triton-lib} \\ (\textbf{this work})\end{tabular} \\ \hline
\multirow{2}{*}{\hspace*{-2ex}\rotatebox[origin=c]{90}{\begin{tabular}[c]{@{}l@{}}\textbf{RMS} \\[-0.5em]\textbf{norm}\end{tabular}}} &
\begin{tabular}[c]{@{}l@{}}\texttt{layernorm\_}\\ \texttt{   kernels.cu}\end{tabular} &
159 &
\begin{tabular}[c]{@{}c@{}}NVIDIA \\ (\& AMD via \\ \texttt{hipify})\end{tabular} &
\href{https://github.com/vllm-project/vllm/}{github.com/vllm-project/vllm}, \cite{KwonEfficientMemoryManagement2023} \\ \cline{2-5} 
&
\textbf{Triton w/ autotuning} &
96 &
AMD / NVIDIA &
\begin{tabular}[c]{@{}l@{}}\href{https://ibm.biz/vllm-ibm-triton-lib}{ibm.biz/vllm-ibm-triton-lib} \\ (\textbf{this work})\end{tabular}
\end{tabular}
	}
	\vspace*{-1.5em}
\end{table}

The \gls{dsl} Triton~\cite{TilletTritonIntermediateLanguage2019} has recently become popular as a promising open-source alternative to writing custom CUDA kernels. Triton (sometimes called \textit{OpenAI Triton}) enables writing and debugging kernels using simple Python code, which can be executed on various GPUs. Triton kernels have been shown to be both highly performant and portable across different GPU platforms. For this reason, Triton is growing in popularity; it is used for many \gls{llm} stacks and is integrated into \texttt{pytorch.compile}. 
Triton leverages a \gls{jit} compiler and builds on the idea of \textit{hierarchical tiles} to automate memory coalescing, shared memory allocation, and synchronization between threads~\cite{TilletTritonIntermediateLanguage2019}. Listing \ref{lst:triton-vector-add} shows a one-dimensional parallelized vector addition in Triton. Triton kernels can be fine-tuned for different workload sizes or target architectures using hyperparameters, also called \textit{kernel configurations}. For example, in Listing~\ref{lst:triton-vector-add}, \texttt{BLOCK\_SIZE} is a configuration parameter that influences the scheduling across the GPU cores.

\section{Study: Can Comprehensive Autotuning Enable LLM Kernel Performance Portability?}

\label{sec:evaluation}
\begin{figure*}[t]
	\vspace*{-1em}
	\centering
	\captionsetup[subfigure]{justification=centering}
	\begin{subfigure}[t]{.5\textwidth}
		\includegraphics[width=\linewidth]{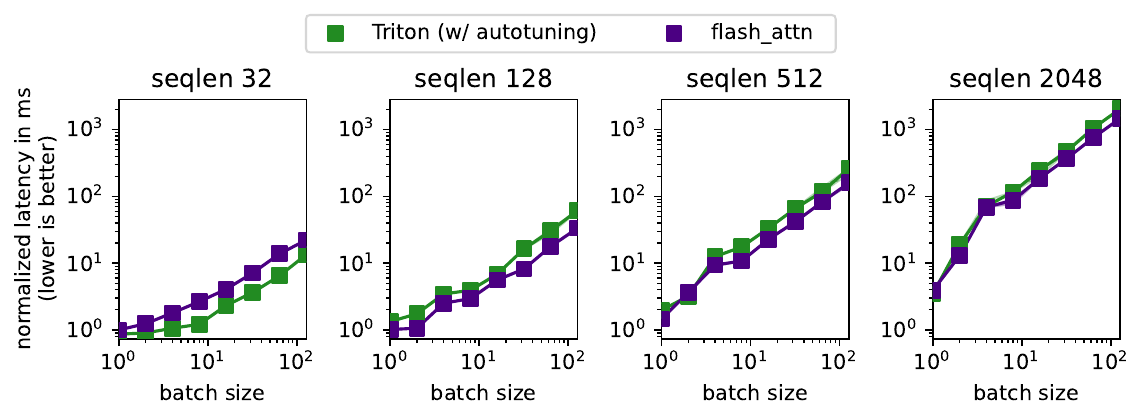}
		\vspace*{-2em}
		\caption{A100}
		\label{subfig:eval-a100}
	\end{subfigure}%
	\begin{subfigure}[t]{.5\textwidth}
		\includegraphics[width=\linewidth]{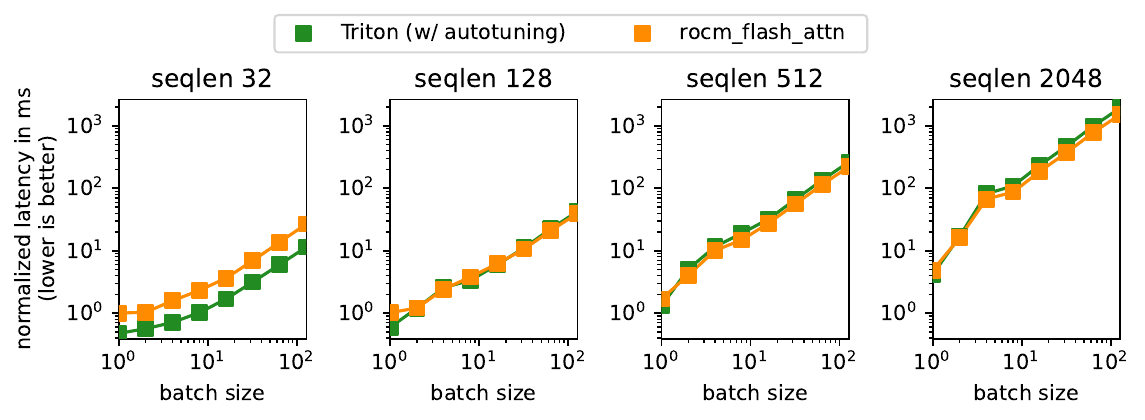}
		\vspace*{-2em}	
		\caption{MI250}
		\label{subfig:eval-mi250}
	\end{subfigure}%
	\vspace*{-0.5em}
	\caption{Comparing performance of causal flash attention implementations.
		\texttt{seqlen} is the maximum input sequence per batch. 
	}
	\label{fig:eval-a100-mi250}
	\vspace*{-1em}
\end{figure*}

\begin{figure*}[t]
	\vspace*{-0.1em}
	\centering
\begin{minipage}[t]{0.2\textwidth}
	\centering
	\hspace*{-0.9em}
	\includegraphics[width=1.08\linewidth]{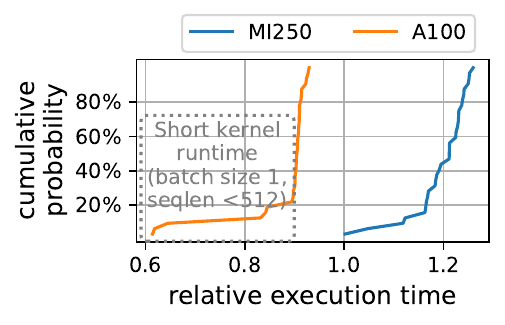}
	\vspace*{-2.5em}
	\captionof{figure}{
		Relative performance of fused RMS norm kernels.
	}
	\label{fig:eval-rms}
\end{minipage}
\begin{minipage}[t]{0.39\textwidth}
	\centering
	\captionsetup[subfigure]{justification=centering}
	\begin{subfigure}[t]{.49\textwidth}
		\includegraphics[width=\linewidth]{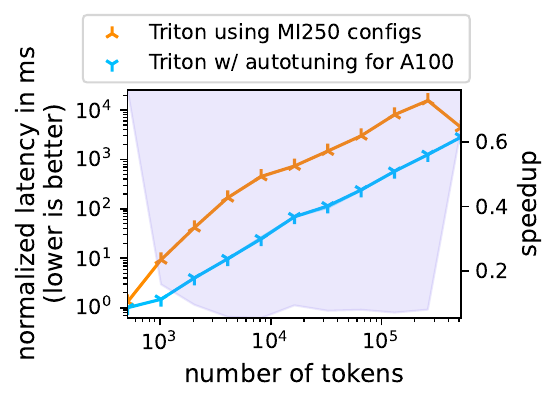}
		\vspace*{-2em}		
		\caption{A100}
		\label{subfig:autotune-baseline-a100}
	\end{subfigure}%
	\begin{subfigure}[t]{.49\textwidth}
		\includegraphics[width=\linewidth]{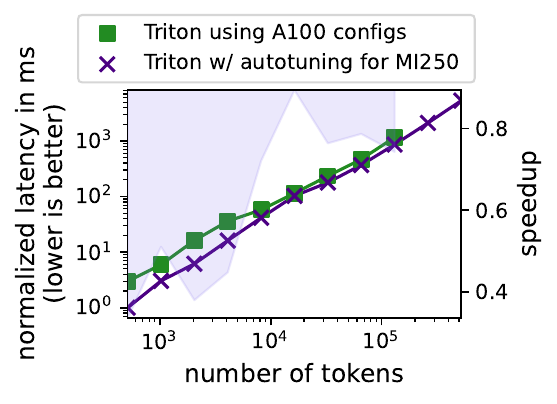}
		\vspace*{-2em}		
		\caption{MI250}
		\label{subfig:autotune-baseline-mi250}
	\end{subfigure}%
	
	\vspace*{-0.5em}
	\captionof{figure}{
		Evaluation of cross GPU configuration portability.
	}
	\label{fig:eval-autotune-baseline}
\end{minipage}
\begin{minipage}[t]{0.39\textwidth}
	\centering
	\captionsetup[subfigure]{justification=centering}
	\begin{subfigure}[t]{.49\textwidth}
		\hspace*{-0.5em}
		\includegraphics[width=1.1\linewidth]{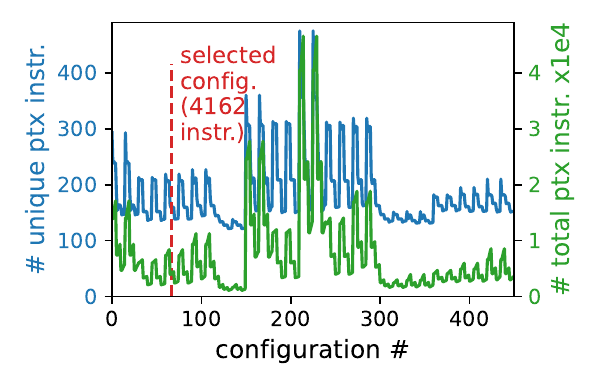}
		\vspace*{-2em}
		\caption{450 Triton configurations}
		\label{subfig:ir-analysis-triton}
	\end{subfigure}
	\begin{subfigure}[t]{.49\textwidth}
		\includegraphics[width=1.1\linewidth]{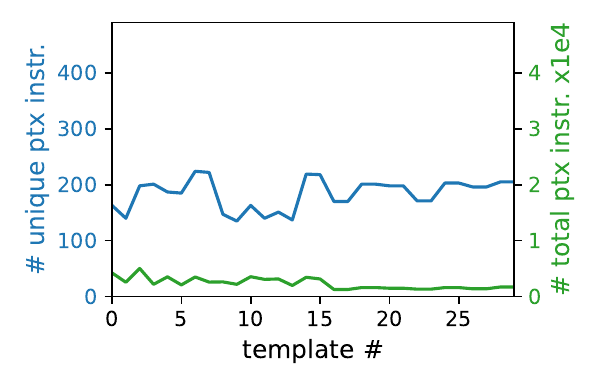}
		\vspace*{-2em}	
		\caption{30 CUDA templates}
		\label{subfig:ir-analysis-cuda}
	\end{subfigure}
	\vspace*{-1.5em}
	\captionof{figure}{
		Diversity of 
		SASS assembly code.
	}
	\label{fig:eval-ir-analysis}
\end{minipage}
\vspace*{-1em}
\vspace*{-0.6em}
\end{figure*}

In this work, we revisit the current state-of-the-art of performance portability for \gls{llm} kernels with a focus on utilizing autotuning.
We make the case for using Autotuning in practice by answering the following research questions:
\begin{enumerate}[label=Q \arabic*)]  
	\item \textit{Can autotuning help achieve LLM kernel portability and \gls{sota} performance?}
	\item \textit{To what extent is autotuning truly necessary?} Would a (\gls{jit}) compiler-only approach be enough?
	\item \textit{What prevents the use of autotuning in today's practice?}
	\item \textit{What is further needed to enable practical autotuning?}
\end{enumerate}

\vspace*{-0.5em}
\subsection*{Method and Investigated Kernels}

We use flash attention kernels as our primary investigation vehicles. Attention is the most performance-critical and complex kernel used by the vast majority of the state-of-the-art \glspl{llm}.
Our flash attention kernel implementation~\cite{DaoFlashAttention2FasterAttention2023} in Triton has \SI{1134}{\loc}, including code for autotuning, and is an improved version of an existing open source kernel~\cite{AMDTritonkernelsteamTriton_flash_attentionpy2024} combined with comprehensive autotuning. 
We further verify our experimental analysis using an additional kernel, the \texttt{RMS layernorm}~\cite{ZhangRootMeanSquare2019}, which is typically the second most computationally expensive and performance critical kernel of today's LLMs.
The details of the investigated kernel implementations are listed in \autoref{tab:impl-used}. All kernels are open-sourced at~\href{https://ibm.biz/vllm-ibm-triton-lib}{ibm.biz/vllm-ibm-triton-lib}. 

We run our evaluation on two GPUs, the NVIDIA A100-80GB and the AMD MI250-128GB. We selected these two GPUs because they utilize comparable technology nodes (MI250 \SI{6}{\nano\meter}, A100 \SI{7}{\nano\meter}), represent two major HW vendors, and also due to their popularity.
We base our kernel parameters on the Llama3-8B LLM architecture (128 head size, 32 query heads, and 8 KV heads) and vary sequence lengths and batch sizes based on real-world samples. The sequences contained within a batch have variable lengths, as it occurs in real-world online inference scenarios. The autotuning time allowed is up to ~\SI{24}{\hour} on each platform, including compilation time, which accounts for around $80\%$ of the autotuning time. 
For data collection, we utilize CUDA/HIP graphs to avoid measuring software-side overheads. 
\vspace{-0.5em}
\questionsec{Can Autotuning enable portable SOTA performance?}

To answer this question, in \autoref{fig:eval-a100-mi250}, we compare the two SOTA implementations of flash attention on A100 and MI250 with our autotuned Triton kernel. Please note, the \texttt{flash\_attn} libraries are \textit{different} for NVIDIA and AMD GPUs, whereas the autotuned Triton kernel is \textit{unchanged}.
\autoref{subfig:eval-a100} and \autoref{subfig:eval-mi250} each show four plots for different maximum sequence lengths. In each plot, the batch size is shown on the x-axis and the latency is denoted on the y-axis (lower values are better). The latency values are normalized by the leftmost latency value of \texttt{flash\_attn}. 

We find that the autotuned Triton kernel is broadly competitive on both platforms, irrespective of batch size or sequence length, while using less than $2\%$ of \gls{loc}. In the best case, the autotuned Triton kernel is up to $2.3\times$ faster than \texttt{flash\_attn}. In the worst case, it still achieves $78\%$ of the SOTA performance -- without any manual optimization.

For the RMS norm, we re-run the same set of benchmarks as in \autoref{fig:eval-a100-mi250}. To execute the CUDA implementation on MI250, it is cross-compiled using \texttt{hipify}, as it is established practice in deployments like vLLM. In the interest of space, we summarize our findings in \autoref{fig:eval-rms} as cumulative distributions that show the relative performance of the autotuned Triton kernel vs. the SOTA baselines. We note a similar trend also for the RMS kernel. The autotuned Triton kernel consistently outperforms the cross-compiled state-of-the-art CUDA code on MI250 by more than $20\%$ on average. For the A100, the autotuned Triton kernel achieves $91-98\%$ in most scenarios, which is promising given that the CUDA implementation was developed and optimized primarily for A100~\cite{KwonEfficientMemoryManagement2023}. However, for small workloads
, the Triton kernel achieves only $60-90\%$ of the A100 baseline. Upon further investigation, we find that the performance difference is due to the Triton compiler not leveraging FP16 optimization opportunities, and is not due to the choice of the kernel parameters that are under the control of the autotuner. 

 
\vspace{-0.5em}
\questionsec{Is autotuning necessary?}
Next, we question if autotuning is a necessary step in achieving SOTA performance portability or if there are simpler alternative paths to achieve the same objective. For example, finding a configuration or a simple heuristic that delivers close to \gls{sota} performance in most scenarios on different accelerators. 
To evaluate whether it is possible to find a good-enough portable configuration, we perform two sets of experiments. First, we select the optimal configuration for each benchmark on each GPU and use this configuration to run on the other GPU. For example, we took the optimal MI250 configuration for a sequence length of 512 and measured it on an A100. 
The results are shown in \autoref{fig:eval-autotune-baseline}. As can be seen, the impact of simply re-using configurations across GPUs is quite dramatic, slowing down execution in the case of using MI250 configurations on the A100 to as little as $7\%$ (\autoref{subfig:autotune-baseline-a100}). Certain configurations from one platform are not even valid on the other platform, as is shown by the missing values at the right-hand side of \autoref{subfig:autotune-baseline-mi250}. Our experiments indicate that performance drops by at least \SI{20}{\%} and by up to an order of magnitude when using a configuration optimized for a different GPU. The high variance of the performance of Triton kernels is also indicated by the large error bars for the manually tuned Triton kernel in \autoref{fig:motivational-example}, where we evaluated five different hyperparameters, equally sampled across the configuration space used in autotuning.

Next, we quantify whether autotuning truly enables better code generation opportunities. Our hypothesis is that autotuning, by exploring different kernel parameters, enables compilation optimizations that the JIT compiler would not be otherwise able to find. To this end, we analyzed the PTX assembly code generated by all $450$ Triton configurations that were evaluated while autotuning one model and one sequence length setup (the Attention layer for LLama3.1-8b, batch size 64, sequence length 2048, see~\autoref{fig:eval-a100-mi250}). 
The analysis is shown in \autoref{fig:eval-ir-analysis}. 
We perform three types of quantitative code analyses: First, we count the number of unique assembly code instructions, evaluating only the opcodes and prefixes without considering the operands. The result is shown on the \textcolor[HTML]{1f77b4}{blue} curve (y-axis on the left). 
Second, we count the total number of PTX instructions in the \texttt{.cubin} file, as shown in the \textcolor[HTML]{2ca02c}{green} curve (y-axis on the right).
%
\autoref{subfig:ir-analysis-triton} shows no relation between these two measures. Also, by looking at both curves in isolation, it is not obvious why configuration \# 67 was chosen as the best configuration by the autotuner, as marked by the \textcolor[HTML]{d62728}{red} marker.
%
We compare the autotuned code with the one generated by the compiler based on CUDA templates. We use all 30 templates applicable to our scenario (micro architecture \texttt{sm80}, data type \texttt{fp16}). 
Comparing \autoref{subfig:ir-analysis-triton} and \autoref{subfig:ir-analysis-cuda} reveals three key differences: 
First, all CUDA library templates use a less diverse set of PTX instructions, as shown by the maximum number of unique PTX instructions (224), which is less than half of what Triton generates (475). Next, the generated \texttt{.cubin} files are not only less diverse in terms of instructions, but they have a smaller and narrower range of sizes. The Triton code generated can be over one order of magnitude larger, indicating that the compiler can introduce code specialization based on techniques such as loop unrolling and software pipelining.  

We limit our code analysis to NVIDIA GPUs, primarily due to space considerations and also because the number of valid Triton configurations for AMD GPUs was significantly lower and therefore less insightful. 
Overall, our code analysis suggests that autotuning facilitates a broader and more efficient exploration of the possible solution space compared to the manually written template libraries.

\vspace{-0.5em}
\questionsec{Why is Triton autotuning not used in practice?}

\label{subsec:autotuning-not-used-triton}

\begin{table}
	\centering
	\vspace*{-1em}
	\caption{Usage of autotuning in popular LLM frameworks}
	\label{tab:autotuner-usage}
	{\notsotiny
		\begin{tabular}{l|r|r|l}
			\textbf{framework} &
			\textbf{\begin{tabular}[c]{@{}l@{}}triton \\ kernels\end{tabular}} &
			\textbf{\begin{tabular}[c]{@{}l@{}}kernels w/ \\ autotuning\end{tabular}} &
			\textbf{source} \\ \hline
			vLLM                    & 57 & 7 & \href{https://github.com/vllm-project/vllm}{github.com/vllm-project/vllm}~\cite{KwonEfficientMemoryManagement2023}     \\
			pytorch-labs/applied-ai & 61 & 9 & \href{https://github.com/pytorch-labs/applied-ai}{github.com/pytorch-labs/applied-ai}                                  \\
			sglang                  & 13 & 0 & \href{https://github.com/sgl-project/sglang/}{github.com/sgl-project/sglang/}~\cite{ZhengSGLangEfficientExecution2024}
		\end{tabular}
	}
	\vspace*{-1.5em}
\end{table}

Outside of the academic literature, autotuning for \gls{llm} applications was also attempted. The most notable example is PyTorch Inductor~\cite{pytorch-inductor-2025}. Inductor is the tuning front-end for \texttt{torch.compile}, and leverages autotuning by simply running different operation implementations sequentially. 
Besides Inductor, there is a built-in autotuner for Triton kernels, which requires a list of potential kernel configurations from the programmer and tries all of them sequentially. 
However, this feature is rarely used in practice. Usually, Triton kernels are hand-optimized for a specific GPU and do not perform equally well on different GPU platforms. 
In \autoref{tab:autotuner-usage}, we summarize a survey of popular \gls{llm} frameworks and inference servers. As can be seen, only a fraction of Triton kernels leverage autotuning. We identify three reasons behind the gap between literature and practice when it comes to autotuning: 

First, using the built-in autotuner adds significant overhead to Triton kernel launches. This overhead stems from the fact that, for every variation in the kernel parameters, the autotuner must determine which kernel version performs best, which requires \gls{jit} compilation and execution.
Additionally, the autotuning process is repeated each time a new process is started, since the autotuner results are only valid within the process that created them. 
As pointed out by previous work (\cite{RingleinRFCAutotunerDejavu2024, MaherCacheAutotuneTimings2025}), these implementation decisions are suboptimal but remain unfixed. 

A second reason why autotuning is not widely used in practice is because performance portability was not the main focus of the \gls{llm} community. Usually, the immediate goal in research and industry is to show performance on a standard set of benchmarks on the \dirquote{de-facto default platform}. Enhancing performance-portability across platforms is a secondary goal. Hence, a lack of code maturity provides another explanation why autotuning is not used in today's practice.

Third, the Triton autotuner still requires some manual guidance. The developer must provide a list of configuration options to explore for each kernel. The list is usually based on the programmer's intuition and has a significant impact on the resulting performance. For example, our results show a difference of nearly $20\times$ for complex kernels (see \autoref{subfig:autotune-baseline-a100}). 


\vspace{-0.5em}
\questionsec{What are the gaps towards practical autotuning?}
\label{subsec:future-dejavu}

We argue that practical autotuning is indeed possible. During our evaluation study, we identified several necessary 
\newpage
\vspace*{-1.5em}  
improvements for making Triton autotuning practical:

\begin{enumerate}[leftmargin=*]
    \item \textbf{Autotuning API}: LLM kernel developers need access to a high-level API to define kernel parameter configuration spaces and also express parameter dependencies.
    
    \item \textbf{Efficient search of the configuration space}: The parameter search space size can be very large. For example, for flash attention, there can be up to ~1000 configurations per tensor shape, some of which are invalid on certain GPU platforms.
    Autotuning needs to leverage advanced search methods 
    to reduce autotuning time and reliably identify optimal configurations.
	
	\item \textbf{Reusable autotuning}: Autotuning results should be cached in a reusable way to avoid unnecessary re-tuning.
    Ideally, autotuning results should contain all relevant environment dependencies to ensure correct reuse and should be stored outside of the LLM deployment.
	
	\item \textbf{Move autotuning off the critical path:} An alternative approach to reducing autotuning overheads is to perform it ahead of time, either as part of the kernel development process or, if not possible, to perform autotuning based on workload metrics using idle GPU times.
\end{enumerate}

\section{Conclusion}

Through this work, we make the case that autotuning is a key component necessary to achieve performance portability for today's \gls{llm} stacks. We focus on attention kernels and show that combining the Triton \gls{jit} compiler with holistic autotuning can enable performance portability across GPUs from vendors. We demonstrate that autotuning explores up to $15\times$ more kernel parameter configurations, produces significantly more diverse code across multiple dimensions, and even outperforms vendor-optimized implementations by up to $2.3\times$, all while reducing kernel code size by $70\times$ and eliminating the need for manual code optimizations. 


 




\begin{spacing}{0.78}
\printbibliography

@online{AMDTritonkernelsteamTriton_flash_attentionpy2024,
  title = {Triton\_flash\_attention.Py},
  author = {{AMD Triton kernels team}},
  date = {2024},
  url = {https://github.com/rasmith/vllm/blob/71f89c5595a6d1c0b9ce29b68344959cb69aa21a/vllm/attention/ops/triton_flash_attention.py},
  urldate = {2025-03-10}
}

@online{DaoFlashAttention2FasterAttention2023,
  title = {{{FlashAttention-2}}: {{Faster Attention}} with {{Better Parallelism}} and {{Work Partitioning}}},
  shorttitle = {{{FlashAttention-2}}},
  author = {Dao, Tri},
  %date = {2023-07-17},
  date = {2023},
  eprint = {2307.08691},
  eprinttype = {arXiv},
  eprintclass = {cs},
  url = {http://arxiv.org/abs/2307.08691},
  urldate = {2024-03-20},
  abstract = {Scaling Transformers to longer sequence lengths has been a major problem in the last several years, promising to improve performance in language modeling and high-resolution image understanding, as well as to unlock new applications in code, audio, and video generation. The attention layer is the main bottleneck in scaling to longer sequences, as its runtime and memory increase quadratically in the sequence length. FlashAttention exploits the asymmetric GPU memory hierarchy to bring significant memory saving (linear instead of quadratic) and runtime speedup (2-4\$\textbackslash times\$ compared to optimized baselines), with no approximation. However, FlashAttention is still not nearly as fast as optimized matrix-multiply (GEMM) operations, reaching only 25-40\textbackslash\% of the theoretical maximum FLOPs/s. We observe that the inefficiency is due to suboptimal work partitioning between different thread blocks and warps on the GPU, causing either low-occupancy or unnecessary shared memory reads/writes. We propose FlashAttention-2, with better work partitioning to address these issues. In particular, we (1) tweak the algorithm to reduce the number of non-matmul FLOPs (2) parallelize the attention computation, even for a single head, across different thread blocks to increase occupancy, and (3) within each thread block, distribute the work between warps to reduce communication through shared memory. These yield around 2\$\textbackslash times\$ speedup compared to FlashAttention, reaching 50-73\textbackslash\% of the theoretical maximum FLOPs/s on A100 and getting close to the efficiency of GEMM operations. We empirically validate that when used end-to-end to train GPT-style models, FlashAttention-2 reaches training speed of up to 225 TFLOPs/s per A100 GPU (72\textbackslash\% model FLOPs utilization).},
  pubstate = {prepublished},
  keywords = {Computer Science - Machine Learning},
  file = {/home/ngl/Zotero/storage/97VZMI4M/Dao - 2023 - FlashAttention-2 Faster Attention with Better Par.pdf;/home/ngl/Zotero/storage/23JUUXDI/2307.html}
}

@online{DaoFlashAttentionFastMemoryEfficient2022,
  title = {{{FlashAttention}}: {{Fast}} and {{Memory-Efficient Exact Attention}} with {{IO-Awareness}}},
  shorttitle = {{{FlashAttention}}},
  author = {Dao, Tri and Fu, Daniel Y. and Ermon, Stefano and Rudra, Atri and Ré, Christopher},
  date = {2022-06-23},
  eprint = {2205.14135},
  eprinttype = {arXiv},
  eprintclass = {cs},
  doi = {10.48550/arXiv.2205.14135},
  url = {http://arxiv.org/abs/2205.14135},
  urldate = {2024-01-24},
  pubstate = {prepublished}
}

@inproceedings{Diamantopoulos2020b,
  title = {Agile Autotuning of a Transprecision Tensor Accelerator Overlay for {{TVM}} Compiler Stack},
  booktitle = {Proceedings of the 30{{{\textsuperscript{th}}}} {{IEEE International Conference}} on {{Field-Programmable Logic}} and {{Applications}} ({{FPL}})},
  author = {Diamantopoulos, D. and Ringlein, B. and Purandare, M. and Singh, G. and Hagleitner, C.},
  %date = {2020-08-31/2020-09-04},
  date = {2020},
  pages = {310--316},
  publisher = {IEEE},
  location = {Gothenburg, Sweden},
  doi = {10.1109/FPL50879.2020.00058},
  %eventdate = {31 August - 4 September 2020},
  isbn = {978-1-7281-9902-3},
  file = {/home/ngl/Zotero/storage/AQ8QIZR7/990200a310.pdf}
}

@article{Hooker2021,
  title = {The Hardware Lottery},
  author = {Hooker, Sara},
  date = {2021-11},
  journaltitle = {Communications of The Acm},
  shortjournal = {Commun. ACM},
  volume = {64},
  number = {12},
  pages = {58--65},
  publisher = {Association for Computing Machinery},
  location = {New York, NY, USA},
  issn = {0001-0782},
  doi = {10.1145/3467017},
  url = {https://doi.org/10.1145/3467017},
  issue_date = {December 2021},
  pagetotal = {8}
}

@online{KwonEfficientMemoryManagement2023,
  title = {Efficient {{Memory Management}} for {{Large Language Model Serving}} with {{PagedAttention}}},
  author = {Kwon, Woosuk and Li, Zhuohan and Zhuang, Siyuan and Sheng, Ying and Zheng, Lianmin and Yu, Cody Hao and Gonzalez, Joseph E. and Zhang, Hao and Stoica, Ion},
  date = {2023-09-12},
  eprint = {2309.06180},
  eprinttype = {arXiv},
  eprintclass = {cs},
  url = {http://arxiv.org/abs/2309.06180},
  urldate = {2024-01-24},
  pubstate = {prepublished}
}

@online{MaherCacheAutotuneTimings2025,
  title = {Cache Autotune Timings to Disk},
  author = {Maher, Bert},
  date = {2025-03-20},
  url = {https://github.com/triton-lang/triton/pull/6261},
  urldate = {2025-03-20},
  organization = {github.com/triton-lang/triton}
}

@article{Moreau2018,
  title = {A {{Hardware}}–{{Software}} Blueprint for Flexible Deep Learning Specialization},
  author = {Moreau, Thierry and Chen, Tianqi and Vega, Luis and Roesch, Jared and Yan, Eddie and Zheng, Lianmin and Fromm, Josh and Jiang, Ziheng and Ceze, Luis and Guestrin, Carlos and Krishnamurthy, Arvind},
  date = {2019-09},
  journaltitle = {IEEE Micro},
  volume = {39},
  number = {5},
  eprint = {1807.04188v3},
  eprinttype = {arXiv},
  eprintclass = {cs.LG},
  pages = {8--16},
  issn = {1937-4143},
  doi = {10.1109/MM.2019.2928962},
  ranking = {rank3}
}

@online{pytorch-inductor-2025,
  title = {{{PyTorch Inductor}} ({{Algorithm Selection}})},
  author = {{PyTorch Community}},
  date = {2025},
  url = {https://github.com/pytorch/pytorch/tree/main/torch/_inductor/select_algorithm.py},
  urldate = {2025-10-03}
}

@article{Ringlein2022,
  title = {Advancing Compilation of {{DNNs}} for {{FPGAs}} Using Operation Set Architectures},
  author = {Ringlein, Burkhard and Abel, Francois and Diamantopoulos, Dionysios and Weiss, Beat and Hagleitner, Christoph and Fey, Dietmar},
  date = {2023-01},
  journaltitle = {IEEE Computer Architecture Letters},
  volume = {22},
  number = {1},
  pages = {9--12},
  issn = {1556-6064},
  doi = {10.1109/LCA.2022.3227643},
  url = {https://ieeexplore.ieee.org/document/9984183/}
}

@online{RingleinRFCAutotunerDejavu2024,
  title = {[{{RFC}}] "Autotuner Deja-vu" Save and Restore Autotuner Cache Persistently},
  author = {Ringlein, Burkhard},
  date = {2024-05-28},
  url = {https://github.com/triton-lang/triton/issues/4020},
  urldate = {2025-03-10},
  organization = {github.com/triton-lang/triton}
}

@online{ShahFlashAttention3FastAccurate2024,
  title = {{{FlashAttention-3}}: {{Fast}} and {{Accurate Attention}} with {{Asynchrony}} and {{Low-precision}}},
  shorttitle = {{{FlashAttention-3}}},
  author = {Shah, Jay and Bikshandi, Ganesh and Zhang, Ying and Thakkar, Vijay and Ramani, Pradeep and Dao, Tri},
  date = {2024-07-12},
  eprint = {2407.08608},
  eprinttype = {arXiv},
  doi = {10.48550/arXiv.2407.08608},
  url = {http://arxiv.org/abs/2407.08608},
  urldate = {2024-10-17},
  pubstate = {prepublished}
}

@inproceedings{TilletTritonIntermediateLanguage2019,
  title = {Triton: An Intermediate Language and Compiler for Tiled Neural Network Computations},
  shorttitle = {Triton},
  booktitle = {Proceedings of the 3rd {{ACM SIGPLAN International Workshop}} on {{Machine Learning}} and {{Programming Languages}}},
  author = {Tillet, Philippe and Kung, H. T. and Cox, David},
  date = {2019-06-22},
  pages = {10--19},
  publisher = {ACM},
  location = {Phoenix AZ USA},
  doi = {10.1145/3315508.3329973},
  url = {https://dl.acm.org/doi/10.1145/3315508.3329973},
  urldate = {2024-01-18},
  eventtitle = {{{PLDI}} '19: 40th {{ACM SIGPLAN Conference}} on {{Programming Language Design}} and {{Implementation}}},
  isbn = {978-1-4503-6719-6},
  langid = {english}
}

@inproceedings{VaswaniAttentionAllYou2017,
  title = {Attention Is {{All}} You {{Need}}},
  booktitle = {Advances in {{Neural Information Processing Systems}}},
  author = {Vaswani, Ashish and Shazeer, Noam and Parmar, Niki and Uszkoreit, Jakob and Jones, Llion and Gomez, Aidan N and Kaiser, Łukasz and Polosukhin, Illia},
  date = {2017},
  volume = {30},
  publisher = {Curran Associates, Inc.},
  url = {https://proceedings.neurips.cc/paper/2017/hash/3f5ee243547dee91fbd053c1c4a845aa-Abstract.html},
  urldate = {2024-01-24}
}

@online{YeFlashInferEfficientCustomizable2025,
  title = {{{FlashInfer}}: {{Efficient}} and {{Customizable Attention Engine}} for {{LLM Inference Serving}}},
  shorttitle = {{{FlashInfer}}},
  author = {Ye, Zihao and Chen, Lequn and Lai, Ruihang and Lin, Wuwei and Zhang, Yineng and Wang, Stephanie and Chen, Tianqi and Kasikci, Baris and Grover, Vinod and Krishnamurthy, Arvind and Ceze, Luis},
  date = {2025-01-02},
  eprint = {2501.01005},
  eprinttype = {arXiv},
  eprintclass = {cs},
  doi = {10.48550/arXiv.2501.01005},
  url = {http://arxiv.org/abs/2501.01005},
  urldate = {2025-01-20},
  pubstate = {prepublished}
}

@online{ZhangRootMeanSquare2019,
  title = {Root {{Mean Square Layer Normalization}}},
  author = {Zhang, Biao and Sennrich, Rico},
  %date = {2019-10-16},
  date = {2019},
  eprint = {1910.07467},
  eprinttype = {arXiv},
  eprintclass = {cs, stat},
  url = {http://arxiv.org/abs/1910.07467},
  urldate = {2024-02-13},
  pubstate = {prepublished},
  keywords = {Computer Science - Computation and Language,Computer Science - Machine Learning,Statistics - Machine Learning},
}

@online{ZhengSGLangEfficientExecution2024,
  title = {{{SGLang}}: {{Efficient Execution}} of {{Structured Language Model Programs}}},
  shorttitle = {{{SGLang}}},
  author = {Zheng, Lianmin and Yin, Liangsheng and Xie, Zhiqiang and Sun, Chuyue and Huang, Jeff and Yu, Cody Hao and Cao, Shiyi and Kozyrakis, Christos and Stoica, Ion and Gonzalez, Joseph E. and Barrett, Clark and Sheng, Ying},
  date = {2024-06-05},
  eprint = {2312.07104},
  eprinttype = {arXiv},
  eprintclass = {cs},
  doi = {10.48550/arXiv.2312.07104},
  url = {http://arxiv.org/abs/2312.07104},
  urldate = {2024-09-16},
  pubstate = {prepublished}
}
\end{spacing}
\end{document}